\documentclass[10pt,leqno]{amsart}
\usepackage{graphicx}
\usepackage{indentfirst,csquotes}

\usepackage{amssymb,amsthm,amsmath}
\usepackage{xcolor,paralist,hyperref,titlesec,fancyhdr,etoolbox}

\titleformat{\section}[display]{\normalfont\huge\bfseries\centering}{\centering\chaptertitlename\thechapter}{10pt}{\Large}
\titlespacing*{\section}{0pt}{0ex}{0ex}
\usepackage{titlesec}
\titleformat{\section}{\Large\bfseries\color{blue}}{\thesection}{1em}{}
\titleformat{\subsection}{\large\bfseries}{\thesubsection}{1em}{}
\hypersetup{ colorlinks=true, linkcolor=black, filecolor=black, urlcolor=black }

\usepackage{amsmath}
\usepackage{graphicx} % Required for inserting images
\usepackage{xcolor}
\usepackage{ulem}
\usepackage{multirow}

\begin{document}

\title{First results of a high sensitivity and transportable  Ring Laser Gyroscope}
\author[Initial Surname]{A. Basti$^{1,2}$, G. Carelli$^1$, G. Di Somma$^2$, A.D.V. Di Virgilio$^2$, \\%}
%\author{
F. Fuso$^1$,
R. Macchia$^3$, E. Maccioni$^{1,2}$, P. Marsili$^1$, and A. Pasqualetti$^3$}
%\maketitle
\date{\today}
%\begin{itemize}
\address{
    $^1$ Department of Physics, University of Pisa, Largo B. Pontecorvo 3, 56127 Pisa, Italy.
    $^2$ Istituto Nazionale di Fisica Nucleare, Sezione di Pisa, Largo B. Pontecorvo 3, 56127 Pisa, Italy
    $^3$ EGO (European Gravitational Observatory, via Edoardo Amaldi 5, 56121 Cascina (PI), Italy.}
%\end{itemize}

\begin{abstract}
Within the GINGER project, aimed at installing an array of large frame ring laser gyroscopes for fundamental physics tests and as part of a geophysics observatory located in the underground laboratory at Gran Sasso, Italy (LNGS-INFN), we are developing a ring laser gyroscope design 
to reduce spurious rotation of instrumental origin and the ability to extend the cavity side length from 1.5 up to 5 m, thanks to the implementation of suitable spacers. The new design led to a prototype, called TRIO, with a side length of 1.52 m, conceived as a transportable instrument. The present paper reports on several preliminary measurements of the Earth angular velocity carried out with TRIO. Results have the twofold objective to assess the instrument performance in the present geometrical configuration and to test validity of the design in view of the GINGER project. To this aim, data obtained with TRIO are compared with typical data acquired with other, previously made, gyroscope prototypes, including the large frame GINGERINO already in operation at the Gran Sasso underground site. 
\end{abstract}
\maketitle
\section{Introduction}
Knowledge of absolute angular velocity is of large importance for fundamental physics, geophysics, and space applications. Among the instruments enabling its accurate measurement,  eminent position have Ring Laser Gyroscopes (RL or RLG), in particular 
Middle or Large Frame Ring Laser Gyroscopes (LF-RLG), with typical perimeters above 6-8 m, have demonstrated a sensitivity in the prad/s range for a measurement duration of 1 s \cite{PRL_sub,PRR, Schreiber2023,G_subdaily}. 
%\textcolor{red}{le referenze vanno risistemate tutte e lo faremo alla fine}

The GINGER project, presently under construction as part of a large underground geophysics observatory (UGGS) at Gran Sasso, Italy  \cite{PRD,MEMOC}, involves an array of LF-RLGs and aims at fully exploiting the excellent sensitivity of the instruments \cite{Schreiber2023,G_subdaily,PRL_sub} for geophysics and geodesy, for instance to measure the variations of the Length of Day (LoD), where the expected improvements are mainly for daily and sub-daily periods. However, measurements will be also relevant for fundamental physics, since the Earth rotation is affected by the de Sitter and Lense-Thirring effects and its investigation can unravel occurrence of Lorentz violations \cite{MEMOC, Tartaglia, Tasson, ML1, A}. 
RLGs are based on a closed path interferometer consisting of a high finesse optical cavity, usually defined by four mirrors at the vertices of a square. An active medium is contained in the square ring cavity and two counter-propagating laser beams are generated \cite{uno}.
Interference of the beams transmitted by each mirror brings information on the non reciprocal effects experienced by the counter-propagating beams. Since the two beams share the same path, differences due to these non reciprocity effects are extremely small. 
However, if the optical cavity is rotating, an asymmetry is generated between the two propagation directions, leading to an effect proportional to the rotational rate. Such  an effect, named after Charles Sagnac and  known since more than a century, dominates all other effects and is  generally exploited in devices for inertial navigation \cite{BigBook, King}. 

The RLG signal is a beat note, whose frequency, called Sagnac frequency $f_s$, is proportional to the scalar product between the ring cavity area vector $\boldsymbol{A}$ and the angular velocity $\boldsymbol{\Omega}$ of the ring itself:
\begin{equation}\label{eq1}
f_s=4\frac{\boldsymbol{A}\cdot \boldsymbol{\Omega}}{P\lambda}\;,
\end{equation}
where $P$ is the perimeter length of the ring cavity and $\lambda$ the wavelength of light.

In the last decade the community involved in developing laser gyroscopes has demonstrated an excellent reliability of the instruments, which can run unattended for months, with sensitivity better than $1$ nrad/s. The related technology is now mature and new applicative fields can be explored. They include, among others,  space astronomy, where the availability of reliable and high-precision measurements of angular quantities is necessary for navigating and pointing purposes.  The X-ray interferometric space telescope XRIstel (to be pronounced kristel), presently under development, will include RLGs onboard \cite{Uttley2021, UTTLEY_SPIE} and open a new discovery field enabling spatially resolved measurement of the coronae of stars tens of light years away, neutron star or black hole-stellar binaries across our galaxy, and supermassive black hole binaries across most of the visible universe. 

A key point for both the envisioned new applications and the further increase of instrument performance, is related to the mechanical design of the RLG. Strict requirements on the overall frame stability have in the past led to the development of monolithic architectures, with a frame made of a single block of ultra-low thermal expansion materials, e.g., zerodur, and all components rigidly attached to the same block. While so-conceived instruments have attained sub-prad/s sensitivity \cite{Schreiber2023}, they exhibit several limitations in terms of cost and integration in complex systems, including those embedded in satellites for astronomy investigations. 

Heterolithic (HL) configurations, obtained by assembling together  the required components, that allows for sufficient flexibility in the integration,  have also demonstrated, with suitable analysis, very high sensitivity \cite{PRL_sub}, showing as well limitations induced by the lack of rigidity of the optical cavity, due to the HL structure.
Two HL RLG prototypes are presently operated by our group: GINGERINO \cite{GING1}, placed in the Gran Sasso underground INFN laboratories, and GP2 \cite{GP2}, located in Pisa INFN laboratories. Few other examples are running in the world \cite{G_subdaily,ROMY,HURST_GIANT,ER1}.

In HL configurations, details of the experimental setup play a big role, since the instrument response depends on the geometry, see Eq.~\ref{eq1}, and spurious rotations due to environmental induced deformations must be necessarily prevented.  Therefore, rigidity and geometrical stability must be ensured by proper mechanical design and appropriate material selection. The use of active controls by remotely actuated mirror translators is also needed to correct residual disturbances, for instance those associated with temperature variations and consequent length changes. 

\begin{figure}[ht]
    \centering
     \includegraphics[scale=0.7]{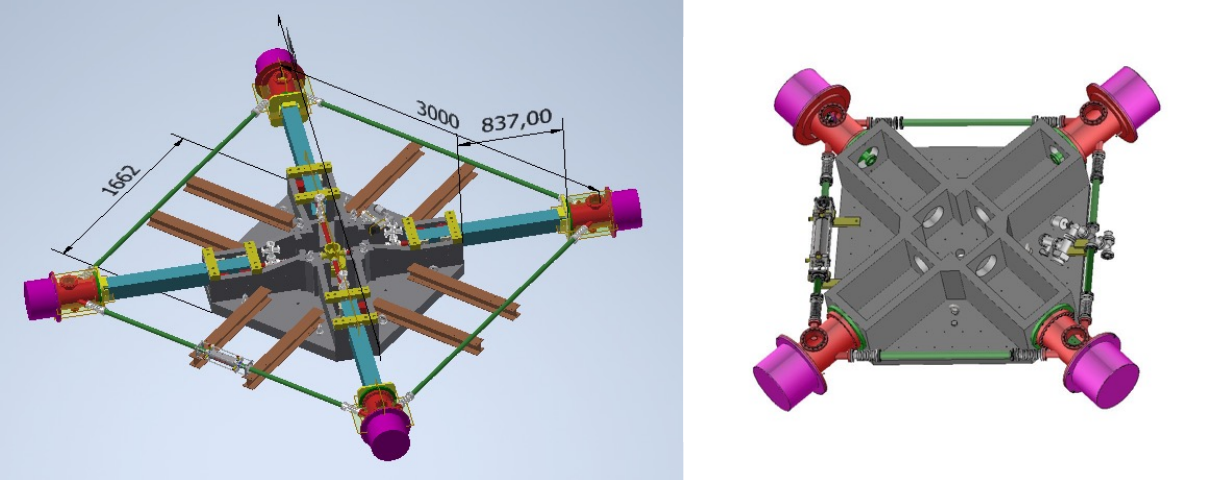}
    \caption{Rendering of the LF-RLG designed for the GINGER experiment (left) and the short-arm version TRIO (right). The central part of the frame, shown in grey and made of granite (side length 1662 mm), is similar by the two RLGs. Suitable spacers 837 mm long, drawn in green in the left panel, allows extending the side of the cavity up to 3000 mm in the shown example. Vacuum tubes are drawn in green; the RF discharge providing the counter-propagating laser beams is included in one side of the square. The vacuum chambers hosting the remotely actuated cavity mirrors are drawn in red; connections between vacuum chambers and tubes are realized through high flexibility bellows to prevent mechanical coupling. Components drawn in magenta represent caps, which can be removed to optically access the output beams. }
    \label{fig:GINGERINOTRIO}
\end{figure}

Based on our past experience, we have designed the GINGER LF-RLG with the purpose to minimize external forces acting directly on the mirrors and prevent mechanical couplings among mirrors. The four mirrors cavity design, chosen for the possibility to use two diagonal light paths for in-situ metrology \cite{Enrico2020}, employs top-quality commercial Ultra High Vacuum (UHV) compatible translators, equipped with remotely controlled actuators (PZT and picomotors, for fine and coarse adjustments, respectively).  A distinctive feature of the designed setup is its modularity that allows to scale the side length from 1.5 m to 5 m, and above, owing to the implementation of suitable spacers, see Fig.~\ref{fig:GINGERINOTRIO}. A short arm version has been built with the main aim to test validity of the design. This led to the availability of a new prototype, called TRIO (Transportable Rotation Interferometry Observatory), that, thanks to its design, is an inherently transportable device. TRIO is presently located in a laboratory shared with other experimental activities, hence, a rather noisy environment, with a $\pm 0.5^o$C thermal stability when doors are kept closed, that allows to assess the instrumental performance in the presence of massive external disturbances.

Objective of the present paper is the first analysis of the TRIO performance through a comparison with data obtained with the GINGERINO and GP2 devices, in order to guarantee that instrumental rotations are reduced with respect to the other prototypes and validate the new HL RLG design for GINGER. The test is as well devoted to TRIO, which later on could be moved to operate in a standard laboratory for geophysical purposes. 
Last but not least, at the best of our knowledge TRIO represents the first demonstration of  a HL RLG prototype whose alignment can be fully operated remotely, hence, minimizing effects of external perturbations on the mirrors.

\section{Experimental setup}\label{setupp}
The TRIO experimental setup is based on a square RLG in which two counter-propagating laser beams circulate inside a closed optical cavity defined by four high-reflectivity dielectric mirrors. The cavity is built on a rigid support structure made of granite (model gnu cr5 from Microplan Group, \texttt{https://microplan-group.com/microplan-italia}), a special granite with a fine grained structure able to reduce temperature induced deformations while ensuring a very good homogeneity. The material has a low linear thermal expansion coefficient, below $6\times10^{-6}$ $^{\circ}$C$^{-1}$, and can be machined with high mechanical precision. Scalability of the structure in view of the GINGER project, see Fig.~\ref{fig:GINGERINOTRIO}, is attained by using spacers designed to be made of silicon carbide.

Mirrors (FiveNine Optics) have a nominal reflectivity of 99.999\%. The small fraction of transmitted light is used for diagnostics and signal reconstruction. In particular, the light extracted from the cavity is sent to a set of photodiodes that acquire radiation intensities along both propagation directions, hereafter denoted as monobeam signals, and, after recombination, the interferometric beat note. 
The optical quality of the mirrors determines the cavity finesse, its ring-down time, and the level of backscattered light, that is particularly relevant in RLGs because it couples the counter-propagating beams eventually leading to systematic artefacts in the reconstructed Sagnac frequency. Mirrors must therefore be treated as both optical and vacuum critical components, since their coating must remain free from dust, organic films, water residues, and other contaminants that could increase absorption and scattering, decrease the cavity finesse, modify the symmetry between the two propagating directions. 

In order to prevent contaminations, the whole cavity is kept in UHV conditions: details on the UHV system, relevant materials and procedures are given in Sec.~\ref{appendix}. 
The mirrors are mounted on precision holders integrated into the vacuum system. Two different vacuum-compatible  mirror holders are used: Newport Optics equipped with pico-motors and Polaris-K1E of ThorLabs, equipped with piezoelectric transducers. The picomotors are voltage driven models with a total travel of about 10 mm and a nominal step size of about 25 nm. They are suitable for alignment procedures and slow corrections of the cavity geometry. Their main role is to bring the cavity into a lasing condition and to optimize the beam path. The piezoelectric transducers (PZT) have a shorter stroke, of the order of a few tens of $\mu$m, and provide finer and faster control. Dedicated vacuum feedthroughs are employed to send control signals to the actuators.

In our previous prototypes, the whole mirror corner assemblies could be rotated and displaced to align the optical cavity. On the contrary, in TRIO the mirrors are moved via remote actuation of their holders. The design principle is that two actuators can be used to control pitch and yaw, whereas three actuators acting coherently can translate a mirror along the diagonal direction. The resulting displacement is relevant for perimeter control, because it changes the cavity perimeter while minimizing unwanted angular perturbations.  

Further to allowing for a smooth remote actuations, the new design enables minimizing mechanical coupling between mirrors and the other components of the RLG, in particular the pipes connecting mirror assemblies and the center of the cavity. All components of the HL configuration are steadily and separately attached to the granite frame. Vacuum tight connection between pipes and mirror assemblies are realized by using soft bellows, as shown in Fig.~\ref{fig:GINGERINOTRIO}. 

The laser medium consists of a He–Ne gas mixture emitting at 632.8 nm, being sustained by a radio-frequency (RF) discharge inside a dielectric tube with a 4 mm inner diameter. A custom-made electrode system is used to ignite the discharge and confine it within a few centimeter length, by controlling the discharge current and therefore the coupling to the RF excitation. %As in our previous prototypes, the discharge is stabilized by adjusting the RF power through a feedback loop using the intensity of one monobeam as the control signal.  
The active medium is a 50:50 mixture of two neon isotopes, $^{20}$Ne and $^{22}$Ne, a standard choice in LF-RLGs to reduce mode competition and favor stable single-mode operation \cite{uno, Giacomelli:gas}.

At the best of our knowledge, TRIO represents the first reported RLG that can be aligned via remote actuation. An indirect assessment of the alignment operation can be found in the measurement of the cavity ring-down, carried out by acquiring the monobeam signal as a function of time after the laser beams have been abruptly switched off. Results of the measurements, carefully accounting for the temporal behavior of the RF discharge, suggest a ring-down time $\tau = 350\pm20$ $\mu$s, a value aligned with the one measured in GINGERINO. 

As already mentioned, the optical readout is based on the weak light transmitted by the cavity mirrors. In particular, the Sagnac signal is obtained by combining the two counter-propagating beams on a photodiode. Considering Eq.~\ref{eq1} and the geometrical parameters of the present cavity, the Sagnac frequency due to Earth rotation is approximately 120 Hz. Photodiode signals are amplified through transimpedance electronics and digitized by the data acquisition system. The beat note is then processed to reconstruct the instantaneous phase and frequency via analytic methods based on the Hilbert transform. 
Monobeam signals are acquired as well and used for both diagnostics and feedback loops, e.g., to control and stabilize the RF discharge power. To this end, the signal from one of the monobeam photodiodes, looking at one propagating beam, is compared with a stable reference voltage. The resulting error signal is sent, after conditioning by a PID circuit, to the RF discharge power controller. As a result, laser power of the selected propagating beam is kept stable in a 0.5\% window. The other monobeam signal is also continuously acquired, since it gives information on gain competition, backscattering, and possible intensity redistributions between the two counter-propagating beams.

 \begin{figure}[ht]
    \centering
    \includegraphics[width=0.75\linewidth]{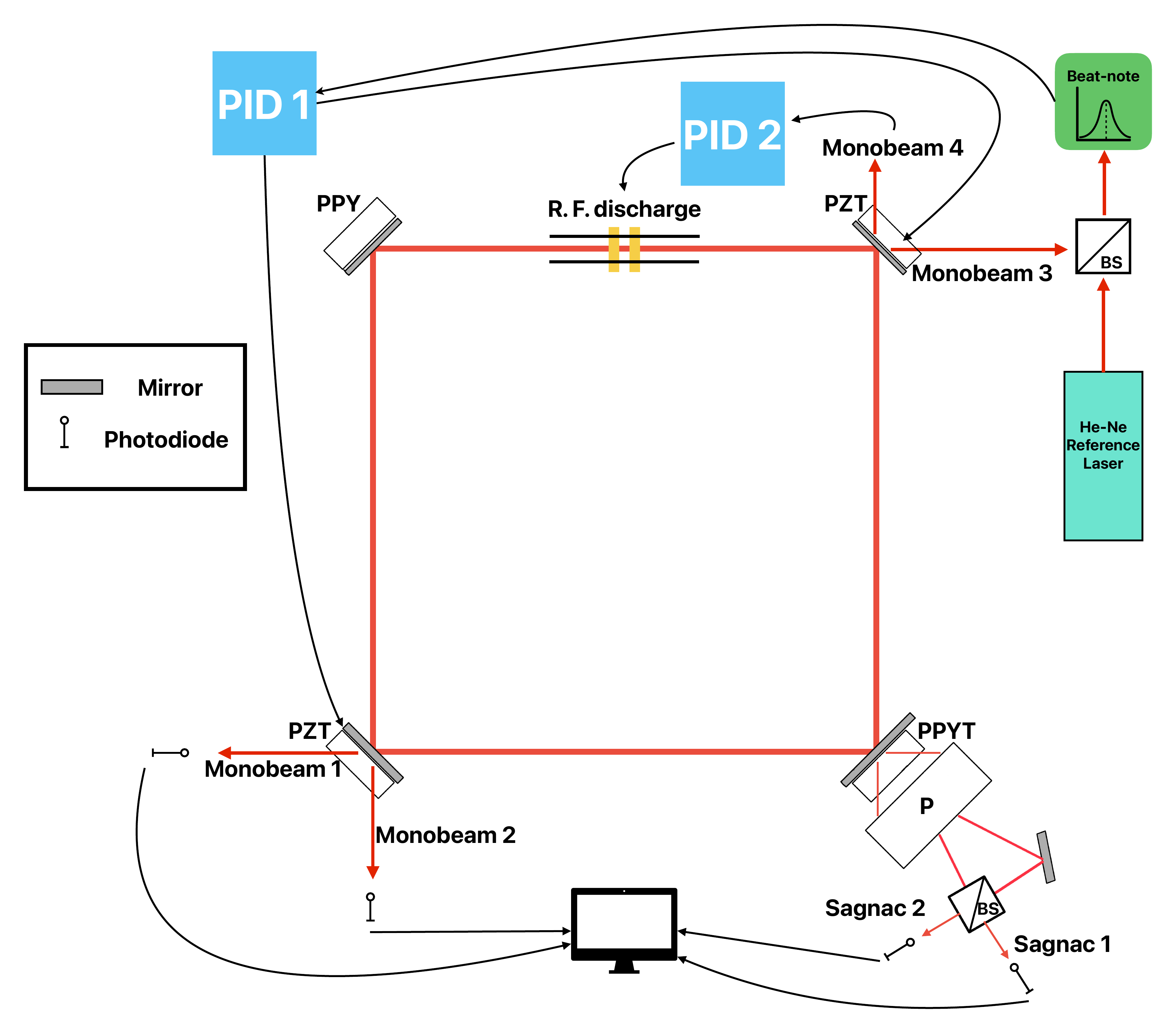}
    \caption{Schematic layout of the cavity setup, with detail of the perimeter control loop. The two counter-propagating beams are optically recombined by the beam splitter located at the lower-right corner, where the Sagnac interferometric signal is generated and detected. The individual beam outputs are monitored at the upper-right corner. On the right side, the perimeter-control loop is shown: one beam is combined with the external He–Ne reference laser, and the resulting beat signal is used in the feedback system that stabilizes the cavity diagonals through the piezoelectric actuators. The abbreviations used in the figure indicate the following components; P: prism; PZT: piezoelectric actuator with pitch, yaw, and translation control; PPY: picomotor actuator with pitch and yaw control; PPYT: picomotor actuator with pitch, yaw, and translation control; BS: Beam Splitter.
}
    \label{fig:set-up}
\end{figure}

 Stabilization of the cavity perimeter is accomplished by a dedicated feedback loop, sketched in Fig.~\ref{fig:set-up}. Briefly, one output beam is combined with an external reference laser (a frequency stabilized He-Ne laser Thorlabs HRS015B), generating a beat note in the range of hundreds of MHz, measurable with a typical  uncertainty of a few hundred Hz. In free-running conditions frequency of the beat note experiences a drift because thermal expansion and contraction, due to the poor thermal control of the room, modify the optical perimeter. When the drift becomes comparable with the optical wavelength, the ring laser may undergo longitudinal mode jumps. Since operation near the edges of the free spectral range enhances competition with adjacent modes, the cavity perimeter is stabilized so that the working point remains near the center of the selected mode interval. To this end, the reference laser beat note is mixed with a second RF signal, stable at the level of about 1 part in $10^9$, and the resulting difference frequency is used as an error signal. Through a PID loop, this signal drives the mirror actuators, which translate the mirrors along the diagonals and stabilize the optical perimeter, suppressing mode jumps over relatively long time intervals. 

 By using the approach, an appreciably stable operation was achieved along the duration of several days, improving the duty cycle, typically $80\%$ for TRIO free running, up to $92\%$. Further refinements are needed to improve the operating stability, reported as $100\%$ in previous applications of a similar approach.
 %with similar technique in the past $100\%$ \cite{Enrico2020} has been obtain, work is in progress to further improve the perimeter control loop.

\section{Results}
Assessment of the TRIO performance has been accomplished by comparing its outcomes with those of the other two prototypes, GP2 and GINGERINO, realized in the past and still in operation in our laboratories. The comparison must account for several design and technical differences, as listed in the following. 

First of all, GP2 and GINGERINO share a similar mechanical design,  but they differ in terms of size (perimeter lengths are about 6.4 m and 14.4 m, respectively), location (in a quiet underground laboratory, in the case of GINGERINO), and orientation. In particular, GP2 is oriented at the maximum Sagnac signal, meaning that the scalar product in Eq.~\ref{eq1} is at its maximum. Conversely, GINGERINO and TRIO have vertical area vectors and measure rotations around such axis. Small tilts in the North-South direction affect at first order the output, since they change the RLG orientation with respect to the Earth rotation axis, whereas GP2 is sensitive only at the second order to North-South tilt, since ƒits area vector is parallel to the Earth axis. 

TRIO, built according to the design outlined in the previous sections, is similar in size to GP2 (side length is 1.52 m vs 1.6 m for GP2). However, contrary to the other two prototypes, which are attached to the underneath basement through reinforced concrete blocks, TRIO, conceived as a transportable instrument, is simply placed on top of a stainless steel optical table. There are also other minor technical differences in the prototype design. For instance, with the aim of reducing overall weight, again to facilitate transportation, UHV components are made in titanium, see Sec.~\ref{appendix}, and the optical arrangement for the readout (lower right corner of the optical cavity in Fig.~\ref{fig:set-up}) includes a prism in front of the output window, required to suitably translate the counter-propagating beams and allow their transmission through the small diameter of the window, mounted in a 63 mm UHV flange.

Despite the expected advantages, both design choices resulted not fully satisfying during the assembly steps. Titanium turned out difficult to be cleaned to the targeted background pressure level (on the order of 10$^{-7}$ mbar), that required prolonged degassing, whereas the output prisms led to increased difficulties in the initial alignment. Therefore, their implementation in the forthcoming construction of GINGER remains under discussion. 

Table \ref{tab:prototypes} summarizes the main design features of the three prototypes. 

\begin{table}[ht]
    \centering
    \begin{tabular}{c|c|c|c|c|c|c}
       \multirow{3}{4em}{prototype} & \multirow{3}{3em}{side length (m)} &\multirow{3}{3em}{area vector}  &\multirow{3}{3em}{bias (Hz)} & \multirow{3}{4em}{remote actuation} & \multirow{3}{4em}{access to diagonals} & \multirow{3}{4em}{reinforced concrete block}\\
       \\
       \\
       \hline\hline
        TRIO & 1.52 & vertical& 120 & YES & YES & NO\\
        \hline
        GP2  &  1.6 &  Earth axis&  184 & NO & YES & YES\\
        \hline
        GINGERINO & 3.6 & vertical & 280 & NO  & NO & YES\\
        \hline\hline
       
    \end{tabular}
    \caption{Summary of the main design features of the RLG prototypes considered in the work. Bias indicates the Sagnac frequency, depending on the geometrical parameters according to Eq.~\ref{eq1}.}
    \label{tab:prototypes}
\end{table}

\subsection{Comparison of sites}

Substantial differences in the operating conditions of the considered RLGs are related to their location, since  site affects relevance and origin of external disturbances. As already stated, GINGERINO runs in an underground laboratory with very small temperature variations (about $0.01^o$C on a daily basis, whereas GP2 and TRIO are installed in standard rooms with worse temperature control within $\pm 0.5^o$C).
Further to temperature variations, seismic noise related to anthropic activities is expected to play a relevant role, since small forces acting on the mirrors can produce small angular rotations of instrumental origin \cite{EPJP_TempVar, ROMY2025}, which are not easy to identify, and in fact limiting the sensitivity of the instrument.
A seismometer analysis, addressing in particular the differences between the underground and standard sites, was carried out. Specifically, the following seismometers data have been used: two Nanometrics Trillium 360 s instruments, until 2 years ago placed on top of the GINGERINO monument and one presently operated at a 80 m distance, as part of the GIGS (Geophysics Interferometer at Gran Sasso) node of the Italian seismic network, and a Guralp CERTIMUS placed in the room hosting TRIO, that is around 20 m apart from GP2. The data of the Trillium located on top of the GINGERINO platform have been removed since it was not performing well at lower frequency, for this reason the one operating 80 m away has been used, after having checked that the two seismometers have very similar behavior in the frequency window where the analysis is focused, mainly below $1$ Hz, in particular at $0.1$ Hz where the minimum of the seismic noise is located.

\begin{figure}[ht]
    \centering
    \includegraphics[scale=0.17]{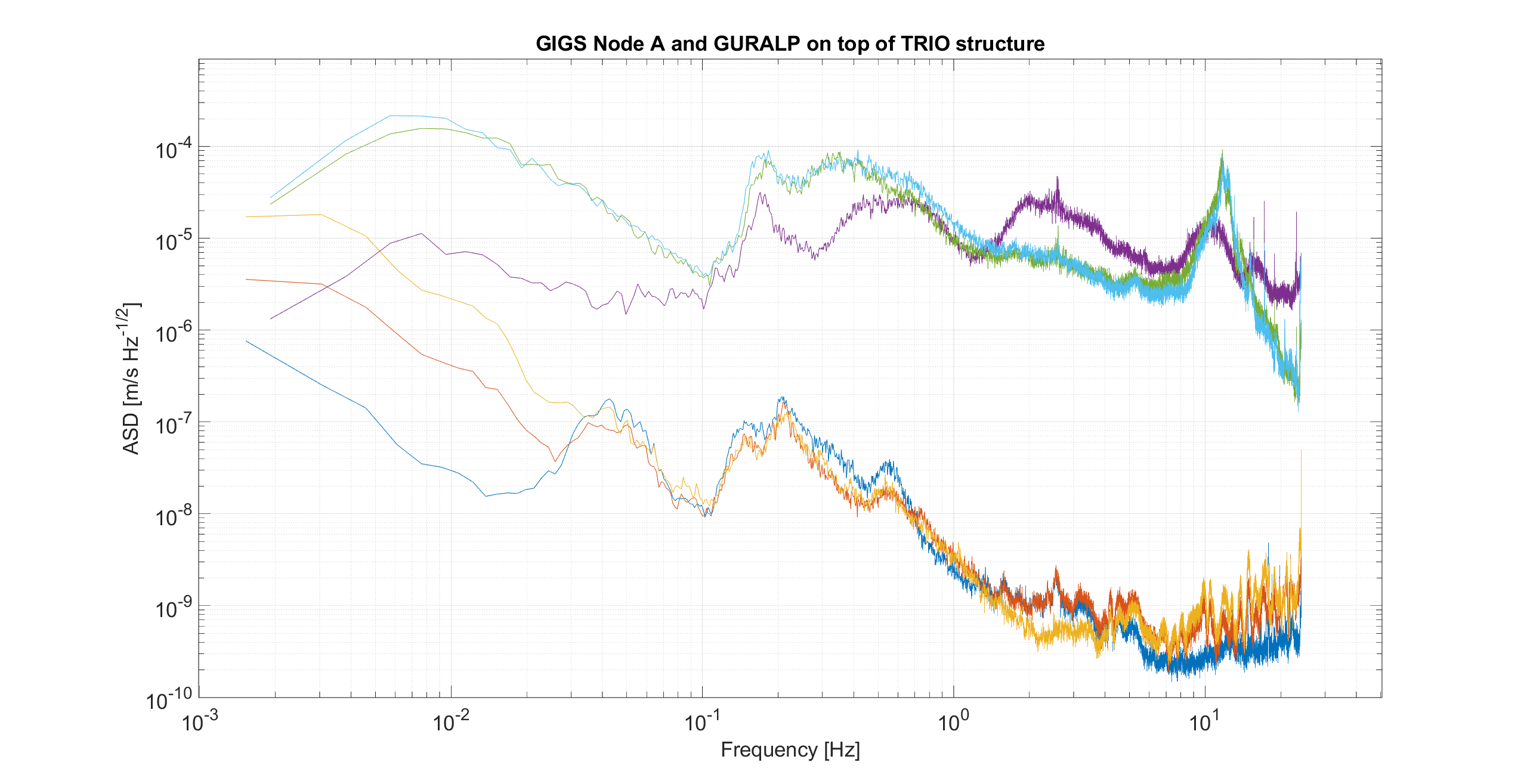}
    \caption{ASD of the acquired seismometer data: top and bottom traces, corresponding to different measurement directions, refer to TRIO and GINGERINO, respectively. We note that, the increase of ASD amplitude in the latter, evident in the frequency range below 20 mHz, is probably due to motion of air in the underground tunnel hosting the RLG.Three hours taken at night have been used for the analysis. Note that the two horizontal translations for TRIO are mainly over-imposed and it is possible to distinguish the two only above 1 Hz. }
    \label{fig:seismometer}
\end{figure}

An example of acquired data is shown in Fig.~\ref{fig:seismometer}, where relevant Amplitude Spectral Distributions (ASDs) are compared. It is clear that the underground location is much more quiet in terms of seismic vibrations, being the ASD amplitude more than two orders of magnitude smaller.

\begin{figure}[ht]
    \centering
    \includegraphics[scale=0.17]{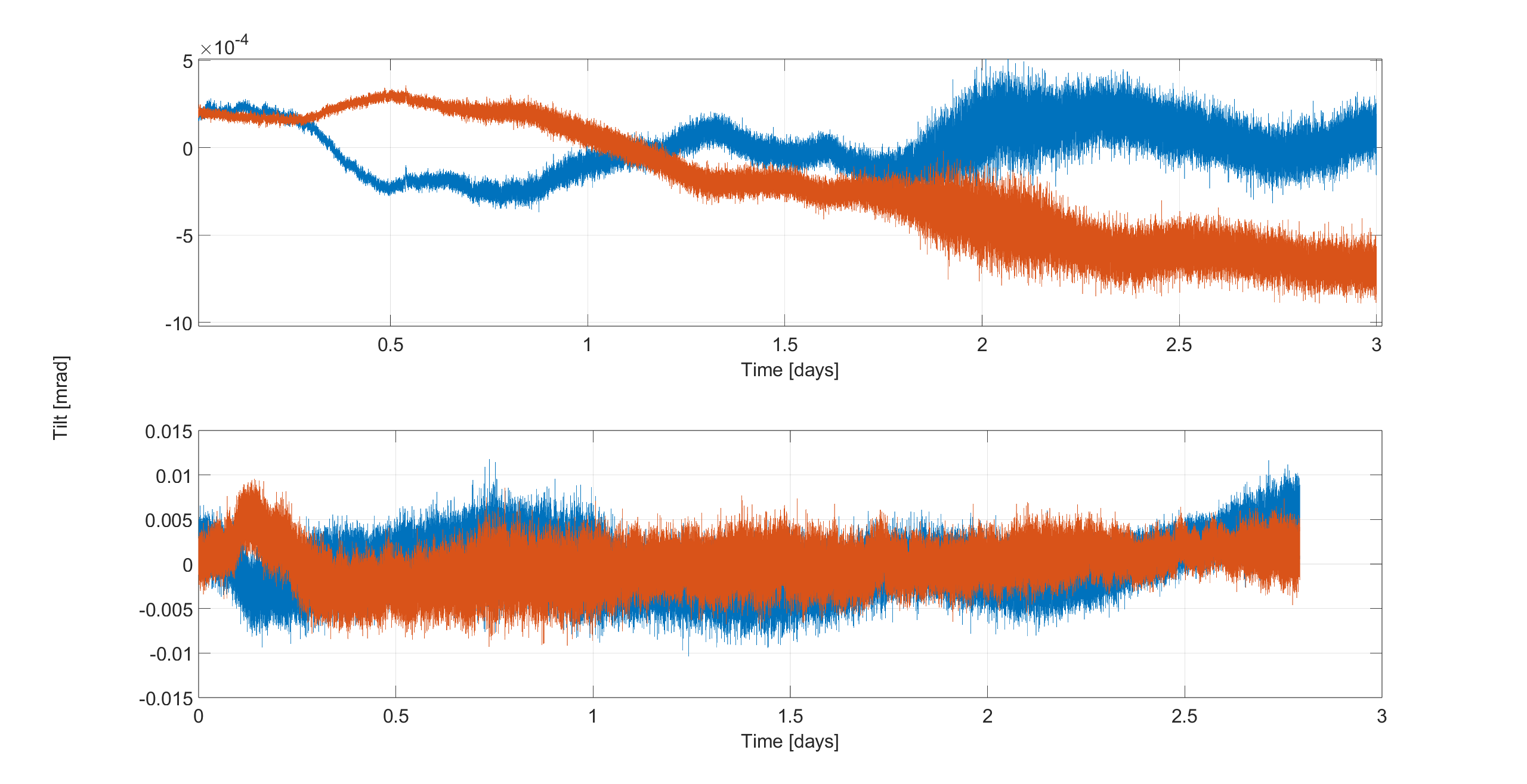}
    \caption{Time series of data acquired by tiltmeters co-located with GINGERINO and TRIO (top and bottom panel, respectively). }
    \label{fig:enter-label}
\end{figure}

An additional analysis was accomplished by using co-located tiltmeters, both 2-K Lippmann 2 components instruments, acquired in a three days interval at 1 Hz rate.  Both tilt amplitudes and relevant fluctuations, evaluated through the standard deviation, are markedly different in the two cases, being typical tilts and standard deviations recorded for TRIO two orders of magnitude, or more, larger than for GINGERINO, as shown in Fig.~\ref{fig:enter-label}.

Seismometer and tiltmeter data unambiguously confirm that, thanks to the smaller noise, an underground location is much more suitable whenever RLGs have to be used for high sensitivity measurements. On the other hand, having TRIO in a noisy site leads to evaluate its performance in non perfect operating conditions, demonstrating the possibility to use such instruments in standard environment, when top sensitivity is not required, as for instance seismic investigation. We note that, since an RLG determines angular velocity based on the frequency reconstruction of the sinusoidal beat-note, its dynamical range is large and the same instrument can virtually be employed for high and low sensitivity measurements \cite{Rubiola}.
%this is in part due to the fact that the dynamic range of RLG is very large, being based on the frequency reconstruction of a sinusoidal signal, and a noisy environment does not require a separate set-up. 

\subsection{Comparison of RLG data}

Comparison of RLG outcomes has been accomplished by using datasets representative of typical operation, acquired in the absence of local or teleseismic earthquakes or other relevant events. For technical motivations, it was in fact impossible to analyze data simultaneously acquired by different instruments. We have applied the procedure developed in the past analysis papers \cite{PRL_sub, PRR, overcoming} to select the best data, i.e., data in which the RLG is in a stationary regime, eliminating all laser mode jump and split mode operations. To this aim, a suitable fringe contrast of the interference, a stable phase mismatch between the two counter-propagating beams, a quantity called $\epsilon$ in our analysis papers \cite{EPJC1, EPJC2}, and the removal of spikes in the reconstructed beat note frequency were required. In the present investigation we use mainly the first two analysis levels, leading to the signals $\omega_m$, the measured beat note, and $\omega_{s0}$, where backscattering contribution has been removed. The third and last level of analysis, $\omega_s$, including also removal of  the laser dynamics effects, is not employed here, being the expected correction small in the considered range of frequencies. It is also worth mentioning that, due to the large Sagnac signal measured by GP2, the relevant correction for the laser dynamics is troublesome and prone to artefacts \cite{EPJC1, EPJC2, PRR}.

\begin{figure}[ht]
    \centering
    \includegraphics[scale=0.17]{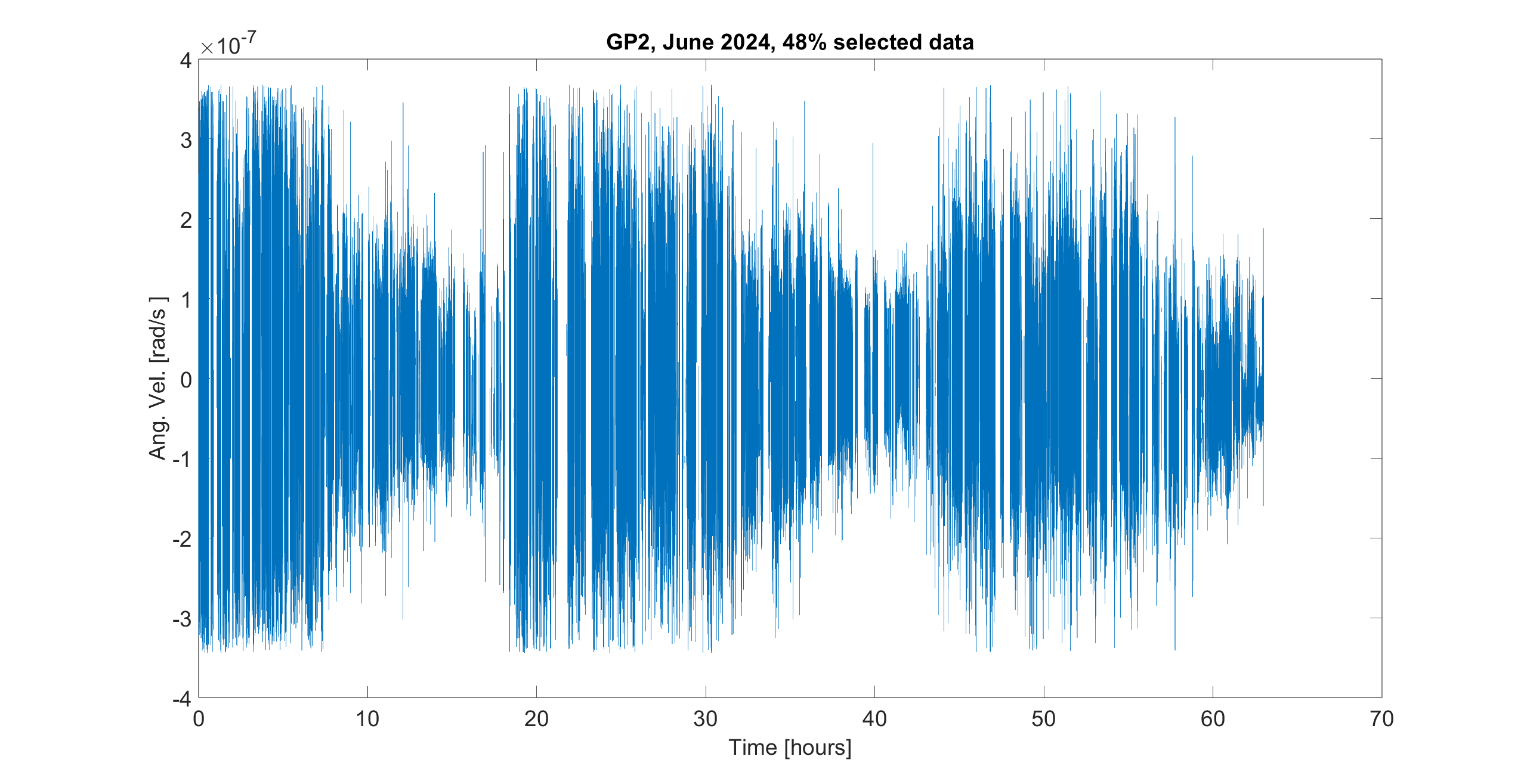}
    \includegraphics[scale=0.17]{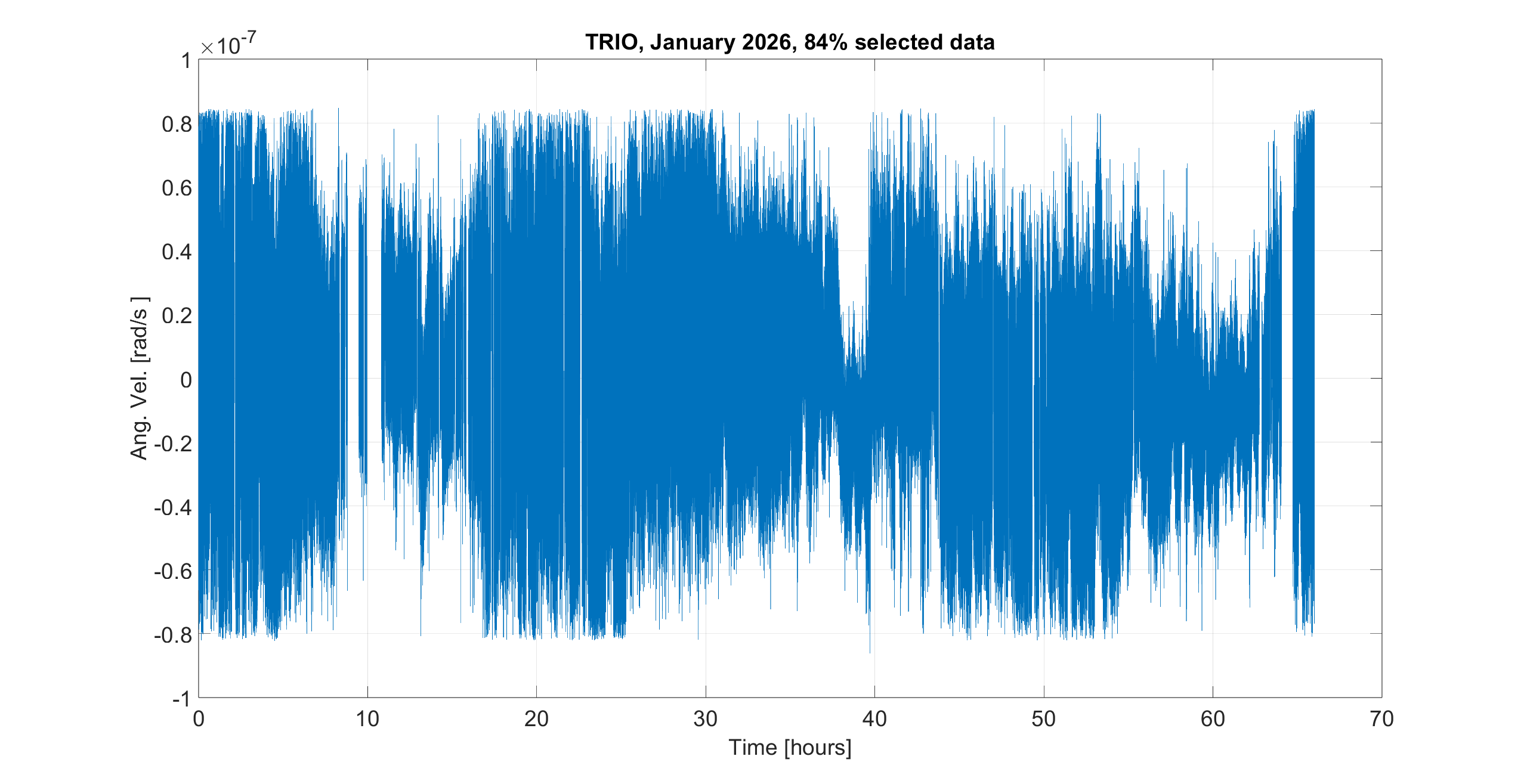}
    \caption{Time series of the angular velocity acquired by GP2 (top) and TRIO (bottom) in a time interval  below three consecutive days. Acquisitions started on June 17, 2024,  and January 23, 2026, for GP2 and TRIO, respectively.}
    \label{fig:timeGP2}
\end{figure}

Approximately three days of the most recent available data of GP2 and TRIO operated in free running conditions, i.e., without perimeter stabilization, have been considered.  Figure \ref{fig:timeGP2} shows the two time series measured in units of angular velocities. The overall data selection ratio is $48\%$ for GP2 and $84\%$ for TRIO,  already suggesting a higher mechanical stability for the latter. 
Moreover, data highlight a smaller peak-to-peak typical variation for TRIO (around 70 nrad/s vs above 150 nrad/s for GP2, as shown in Fig. \ref{fig:timeGP2}).  In both time series, a day/night dependence is  observed, likely due to periodical modifications of temperature and noise, that, however, is more marked for GP2.

\begin{figure}[ht]
    \centering
    \includegraphics[scale=0.17]{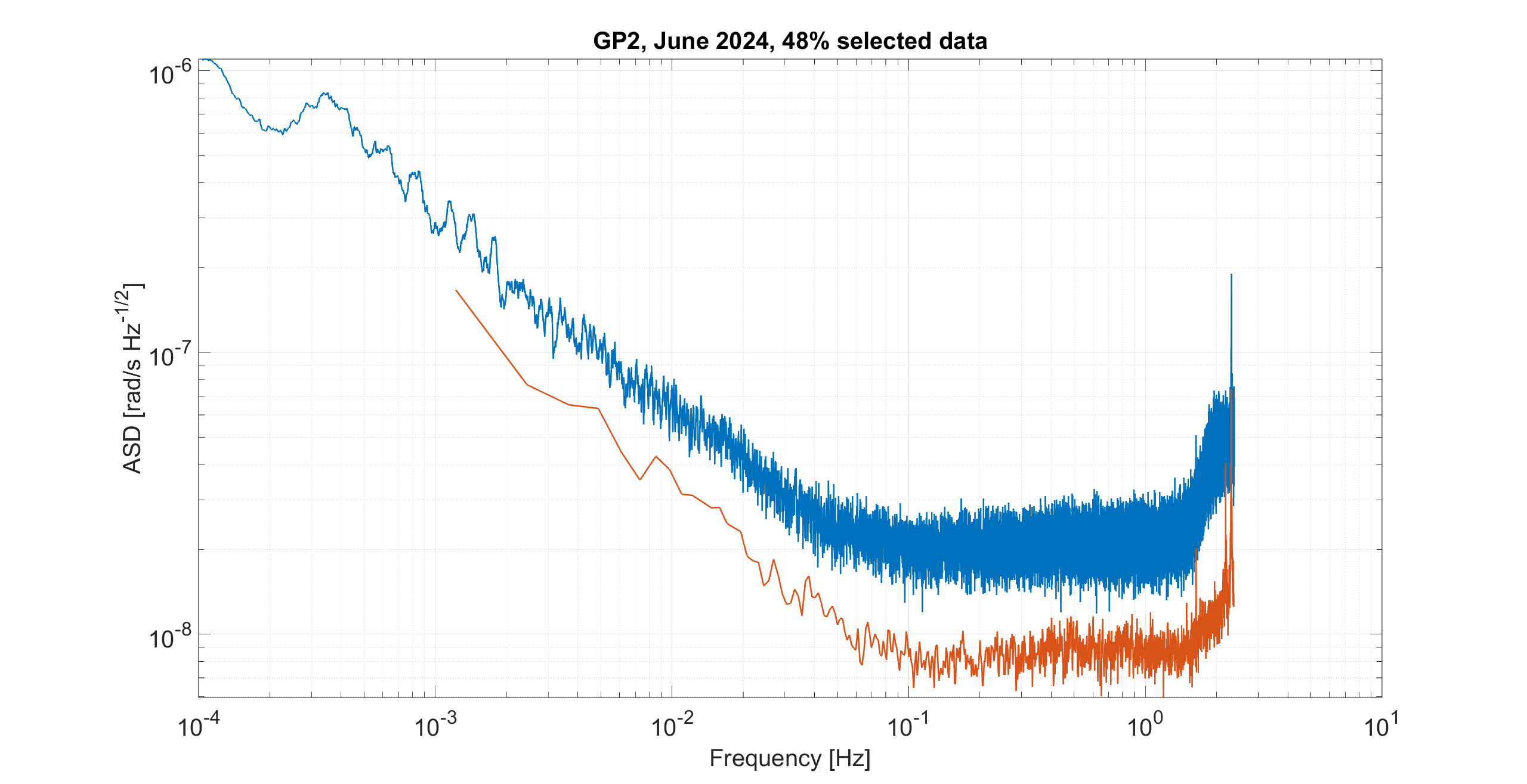}
    \includegraphics[scale=0.17]{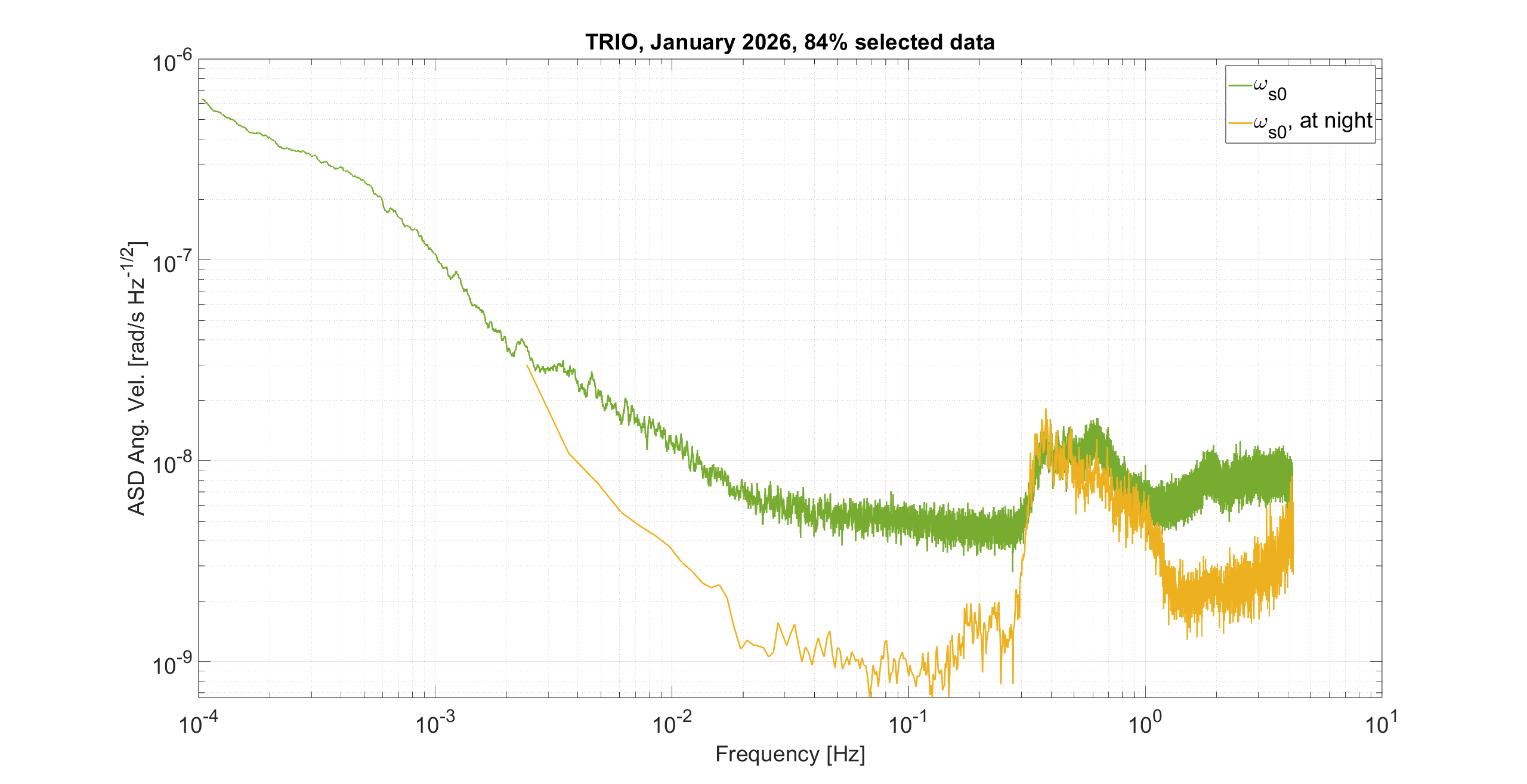}
    \caption{ASD of the time series shown in Fig.~\ref{fig:timeGP2} for GP2 and TRIO (top and bottom panel, respectively). Distinct traces are plotted referred to day and night periods (blue and red on top, and green and amber on bottom panel, respectively).}
    \label{fig:ASD_GP2}
\end{figure}

Results were analyzed by calculating the relevant ASD, as in Fig.~\ref{fig:ASD_GP2}, where data corresponding to the day and night periods are plotted separately. TRIO shows a consistently smaller ASD level when compared to GP2, despite, being both located at a small distance with each other, a similar amount of seismic noise is expected.

%For GP2 the minimum of the Amplitude Power Spectrum (ASD) remains close to $10$ nrad/s$\sqrt{Hz}$, see red curve in  Fig.\ref{fig:ASD_GP2}. 

\begin{figure}[ht]
    \centering
    \includegraphics[scale=0.17]{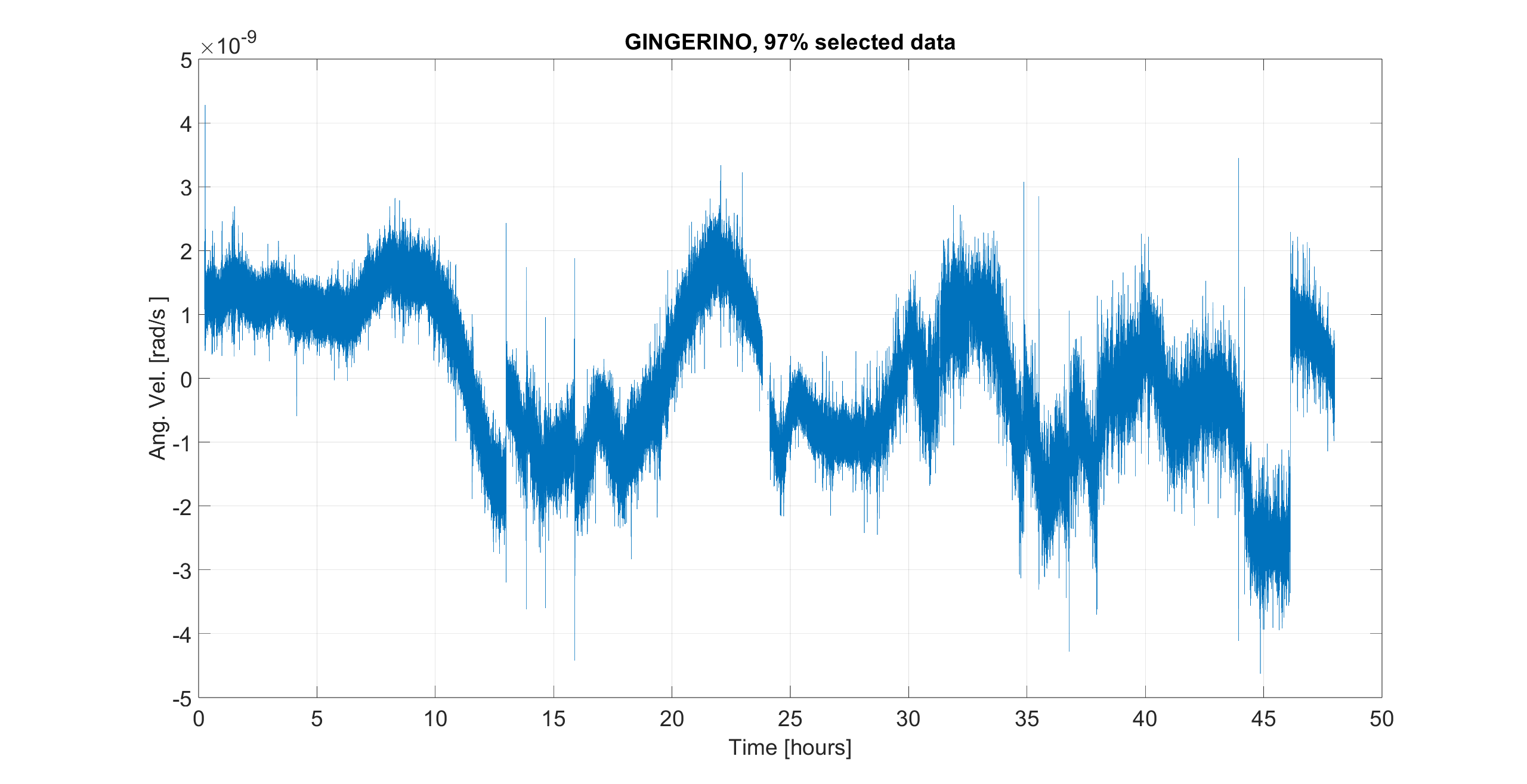}
    \caption{Typical time series of the angular velocity acquired by GINGERINO in a time interval  of around two consecutive days.}
    \label{fig:timeGING}
\end{figure}

A typical two day time series for GINGERINO is shown in Fig.~\ref{fig:timeGING}. Thanks to the much more controlled operating conditions enabled by the underground location, along with the increased sensitivity due to the larger size of the RLG, data dynamics is consistently smaller than for TRIO and GP2, as evidenced by the vertical axis range. Moreover, quietness of the location is also reflected in the very large selected data ratio, $97\%$ in the shown example.

\begin{figure}[ht]
    \centering
    \includegraphics[scale=0.17]{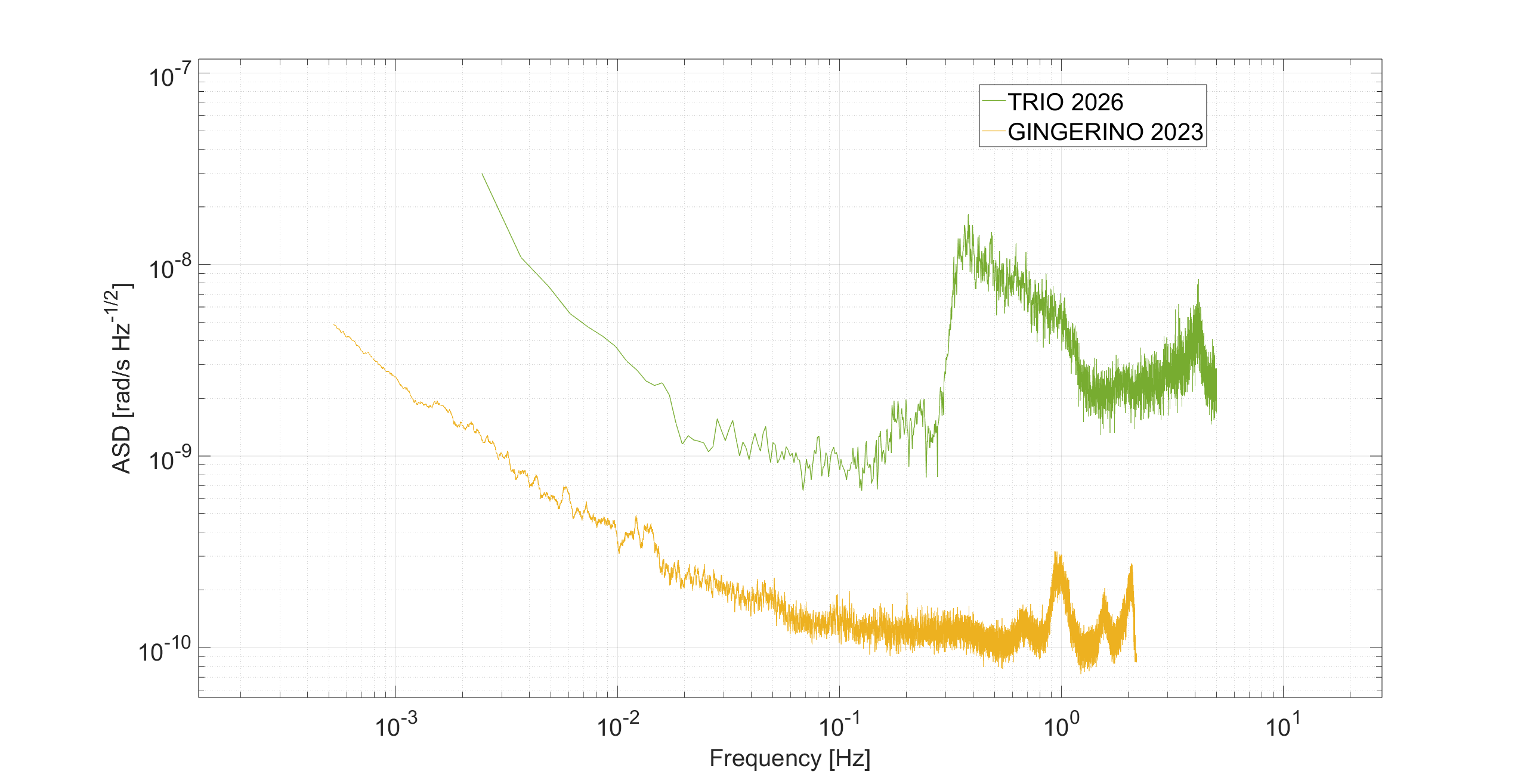}
    \caption{Comparison between the ASD of GINGERINO and TRIO (same data of Fig.~\ref{fig:ASD_GP2}, night period). We note that the bump in the TRIO ASD amplitude around 0.4 Hz can be possibly related to the microseismic noise typical of the site.
}
    \label{fig:2hour}
\end{figure}

The ASD of GINGERINO and TRIO are compared with each other in Fig.~\ref{fig:2hour}. We note that, in the frequency range below 1 Hz, the typical ASD amplitude of TRIO is approximately one order of magnitude larger than for GINGERINO. Further to the different operating conditions, comparison must also account for the different scale factor due to the RLG size. This leads to TRIO having an expected sensitivity smaller by a factor 0.42. Therefore, we can assume that the typical ASD amplitude of TRIO is just a factor 4 larger than GINGERINO, despite the much more noisy environment.

Table \ref{tab:prototypes1} summarizes some of the main results of the analysis.

\begin{table}[ht]
    \centering
%      \begin{tabular}{c|c|c|c|c|c|c}
%       \multirow{3}{4em}{prototype} & \multirow{3}{2em}{data sel. (\%)} &\multirow{3}{3em}{std (nrad/s)}  &\multirow{3}{4em}{monobeam diff. (\%)} & \multirow{3}{7em}{ASD at 0.1 Hz (nrad/s Hz$^{-1/2}$)} & \multirow{3}{2em}{$\tau$ ($\mu$s)} & \multirow{3}{2em}{LD} \\
%       \\
%       \\
%       \hline
%       \hline
%        TRIO & 84 &  $1.8$ & $50$ & 0.9 & 350 &YES\\
%        \hline
%        GP2  &  50-60 &  $5.5$ & $73$ & 8  & 100 & NO\\
%        \hline
%        GINGERINO & 80  & $9.8$ & $49$  & 0.1  & 200 &YES\\
%        \hline \hline
%       
%    \end{tabular}
%    
     \begin{tabular}{c|c|c|c}
       prototype & data sel. (\%) &std (nrad/s) & ASD at 0.1 Hz (nrad/s Hz$^{-1/2}$) \\
       \hline
       \hline
        TRIO & 84 &  $1.8$ & 0.9 \\
        \hline
        GP2  &  50-60 &  $5.5$  & 8 \\
        \hline
        GINGERINO & 80  & $0.98$ & 0.1 \\
        \hline \hline
       
    \end{tabular}
    \caption{Summary of some of the main results of the comparison between different RLG prototypes. Typical data selection ratios are reported in the second column, while std in the third column indicates typical standard deviation of the measured angular velocity. The ASD amplitude in the fourth column reports typical data at 0.1 Hz.}
    \label{tab:prototypes1}
\end{table}
%Important parameter of optical cavity is the ring--down time of the cavity, called $\tau$, it is measured using a fast photomultipler and observing the decay of the transmitted light, when the discharge is abruptly turned off, care is taken in order to eliminate the first part of the signal affected by the discharge itself; in the case of TRIO $\tau = 350 \pm 20 \mu s$ has been measured; the measured $\tau$ is compatible with the losses expected taking into account the mirrors losses, this figure guarantees the feasibility of a good cavity alignment acting remotely on the actuators under vacuum located. 

\begin{figure}[ht]
    \centering
    \includegraphics[scale=0.8]{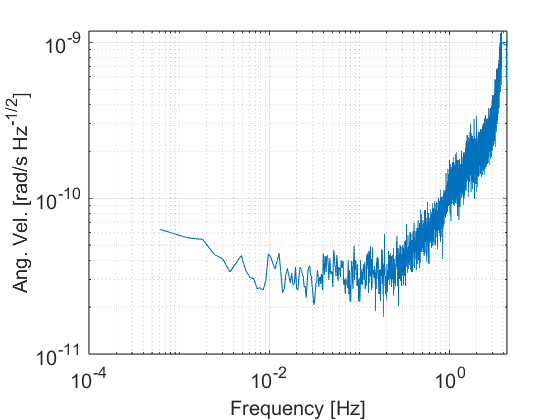}
    \caption{
  Evaluated upper limit for the stochastic noise of TRIO using the difference of the two Sagnac signal exiting the beam-splitter.} 
    \label{fig:RES}
\end{figure}

% RLG has the possibility to measure the limiting noise upper limit\cite{PRL_sub}. The approach is based on the possibility to simultaneously reconstruct two independent beat notes as enabled by observing the two output ports of the cube beam splitter included in the readout setup, see bottom right corner in Fig.~\ref{fig:set-up}. Calling $\omega_{1s0}$ and $\omega_{2s0}$ such two reconstructed beat notes, their difference, $\omega_{res}= \omega_{1s0} - \omega_{2s0}$, is expected to suppress contribution of common mode noise, mostly associated with external disturbances, while highlighting the noise of the measurement itself, and inherently related to the instrument behavior. Figure \ref{fig:RES} shows this limiting value for TRIO, it is relevant to say that this limit is close to the requirement for space application of $10prad/s$ in 1 second measurement. 
 It is relevant to underline how results of the ASD analysis are related to the instrumental sensitivity, but they do not represent a direct information on the limiting noise of the apparatus owing to the presence of external disturbances. Evaluation of the latter can be accomplished based on a recently developed approach \cite{PRL_sub, ADV-2026}. Briefly, by observing the two output ports of the beam splitter shown at the right bottom of Fig.~\ref{fig:set-up}, two independent beat notes, Sagnac 1 and Sagnac 2 in the figure, can be analyzed. Calling $\omega_{1s0}$ and $\omega_{2s0}$ the corresponding reconstructed beat note frequencies, their difference $\omega_{res} = \omega_{1s0}-\omega_{2s0}$ is expected to suppress contribution of the common mode noise, mostly associated with the external disturbances, while extracting the noise of the instrument inherently related to its behavior. As shown in Fig.~\ref{fig:RES}, the upper limit of the instrumental noise is suggested to show a minimum in the frequency interval 10$^{-2}$-10$^{-1}$ Hz. Remarkably, the estimated noise at the frequency of 1 Hz, relevant for space applications, is around 0.1 nrad/s Hz$^{-1/2}$, close to the requirements of the application.
%Figure 9: Evaluated upper limit for the shot noise using the difference of the two Sagnac signal exiting the beam-splitter.

% It is important to distinguish between the minimum signal effectively measured, as shown in the spectral density, and the limiting noise of the apparatus, Fig. \ref{fig:2hour} indicates that the minimum is at $0.1$Hz and TRIO is close to 1 nrad/s in 1 second measurement. 

\section{Conclusions}
TRIO, a new RLG prototype, has been realized according to the design developed within the GINGER project. The design, based on a HL architecture that can be scaled in size, shows several distinctive features aimed at reducing rotations of an instrumental origin induced by environmental changes, a common drawback of HL configurations. Comparison of typical Earth rotational data acquired with TRIO and the other prototypes demonstrates the validity of the design, in particular the use of commercial mirror holders equipped with remote actuations. Further to enabling feedback-based cavity geometry control, that was preliminarily assessed for perimeter stabilization, the remote actuation allowed by PZT and picomotors turned out perfectly suitable for the purposes of optical alignment,  as suggested by the measured ringdown time $\tau$, closed to the expected one based on the quality of the mirrors. Comparison with the GP2 prototype was advantageous for TRIO in terms of all considered parameters such as, duty cycle,  sensitivity, day/night difference. Comparison with GINGERINO was heavily affected by the noisy location of TRIO, established via seismometer and tiltmeter measurements. Nonetheless, results, evaluated by ASD and accounting for the different sensitivity due to the smaller perimeter, suggest that TRIO achieved a performance only a factor 4 worse than GINGERINO, despite the occurrence of a seismic noise larger by more than two orders of magnitude.

Owing to its relatively small size and weight, TRIO can be considered a transportable instrument. Within this frame, it showed a sensitivity already adequate for, e.g., seismic monitoring campaigns. Moreover, the performance was able to meet the general requirements for new envisioned space applications of RLGs, which can also take advantage from the modular design. 

The first assessment of TRIO has been completed, providing reliable validation of the design in view of the challenging objectives of the GINGER project. Thanks to our preliminary tests, several details of the design resulted not fully compliant with the planned goals, for instance, the choice of titanium as the material for the UHV components, leading to serious degassing problems, or the planned size of the output viewports, imposing the use of output prisms leading to a cumbersome initial alignment of the optical cavity.

Further steps will involve a long term investigation of the instrument behavior and the improvement of the geometry control procedures in order to increase the long term stability of the apparatus.

\section{APPENDIX: UHV SYSTEM}\label{appendix}
In the attempt to attain a relatively low weight RLG, conducive to its transportation as well as to space applications, titanium was chosen as the material for the UHV system and its components.
High-vacuum preparation is a fundamental requirement for stable operation of TRIO. The laser cavity contains a gas mixture and optical surfaces with extremely low losses. The presence of residual contaminants can affect the discharge, modify the gas composition, increase mirror losses, increase backscattering effects, and introduce time-dependent changes in the laser dynamics.

The vacuum system must therefore be prepared with two simultaneous goals. The first one is to obtain a sufficiently low base pressure before gas filling. The second one is to minimize the release of contaminants during RF discharge operation. This aspect is particularly relevant because contamination of the cavity does not only affect the vacuum level, but can also modify the standard operating parameters of the RF plasma. Residual gases and organic contaminants can change the discharge impedance, the plasma composition, the excitation balance of the He--Ne mixture, and the effective gain conditions. In severe cases, the laser may fail to reach threshold or may operate in an unstable regime.

This issue was observed during the first tests performed with the titanium system when the cleaning procedure had not yet been fully optimized. The RF discharge did not show the usual orange colour expected for a clean He--Ne plasma, but appeared shifted toward a pink emission. This was interpreted as an indication of contamination or of an altered gas composition inside the cavity. For this reason, dedicated optical emission spectroscopy measurements were performed on the discharge, together with mass-spectrometry analyses of the residual gas, in order to identify the residual species and assess the origin of the contamination.

In the case of the first TRIO titanium cavity, which will not be adopted again for future implementations, an additional cleaning campaign was carried out with the support of the Virgo team. This was necessary because the dimensions of the mechanical parts required dedicated ultrasonic cleaning baths and baking ovens suitable for large components used in UHV environments. The Virgo infrastructure provided the possibility to treat large vacuum components with procedures more appropriate for low-contamination operation than those available in a standard mechanical workshop.

The experience gained with the titanium cavity showed that the cleaning procedure cannot be considered an auxiliary operation, but must be included as a design requirement of the apparatus. For the future GINGER implementation, stainless steel is preferred not only for mechanical and vacuum robustness, but also because its preparation for high-vacuum and ultra-high-vacuum applications is more standardised. The final component procurement should therefore include explicit cleaning specifications, material certificates, and, where possible, documentation of the electropolishing, ultrasonic washing, and baking steps performed before delivery.

Hydrocarbon contamination is a known risk in RLGs operated in UHV. Even small organic residues can become problematic when exposed to RF plasma. Plasma fragmentation can produce deposits on nearby surfaces and can gradually modify the optical and electrical properties of the cavity. In particular, hydrocarbon films on mirrors or viewports can increase optical losses and scattering.

The standard procedure for attaining the base pressure in pre-treated (cleaned) chambers is as follows. After assembly, the system is pumped down to remove air, water vapor, and volatile contaminants. A staged pump-down is preferable to avoid mechanical stress and to allow diagnostic monitoring of the pressure evolution. Bake-out may be required to reduce adsorbed water and slow outgassing species, but the allowed temperature is constrained by mirrors, PZTs, picomotors, feedthroughs, seals, and any internal cables or adhesives.

Residual gas analysis (RGA) should be performed before gas filling. The dominant species should be identified, and particular attention should be paid to water, oxygen, nitrogen, hydrogen, and hydrocarbons. The system is filled with the He--Ne gas mixture only when the residual gas composition is compatible with stable discharge operation. A final base pressure of approximately $3\times10^{-7}$ mbar is typically reached before filling. The cavity is then filled with an isotopic neon mixture at a partial pressure of about $0.25$ mbar, followed by helium up to a total pressure of approximately $8$ mbar.

A SAES GETTERS CapaciTorr Z200 getter pump is permanently operated on the vacuum system and plays a crucial role in removing hydrogen released both by the internal surfaces and by the discharge plasma during operation.

\section*{ACKNOWLEDGMENTS}
This work has been supported by Gaetano De Luca, National Institute of Geophysics and Vulcanology. We acknowledge the big support of Nicolò Beverini, University of Pisa, who passed away one year ago. We have to acknowledge the support and advice of Gabriele Demagistri,  AD and President of Microplan, in the design and realization of the GINGER and TRIO mechanical design.

%The TRIO construction has been supported in part by the STRIC+ project.}

 FINANCED UNDER THE NOTICE FOR THE SELECTION OF PROJECTS FOR THE PROMOTION OF RESEARCH, TECHNOLOGY TRANSFER, AND UNIVERSITY EDUCATION TO BE FINANCED IN THE REGIONS OF LAZIO, ABRUZZO, UMBRIA, AND MARCHEAFFECTED BY THE 2016 SEISMIC EVENTS” – DDG ACT No. 283 OF 2021 – PRESIDENCY OF THE COUNCIL OF MINISTERS – DEPARTMENT FOR COHESION POLICIES – USING THE RESOURCES REFERRED TO IN ART. 1, PARAGRAPH 194 OF LAW NO. 178 OF 30 DECEMBER 2020 - GU GENERAL SERIES No. 322 OF 30-12-2020 ORDINARY SUPPLEMENT No. 46”. CATEGORY A: CREATION OR ENHANCEMENT OF RESEARCH INFRASTRUCTURES PROVIDED FOR UNDER MEASURE B, SUB-MEASURE B.4 (PNRR SUPPLEMENTARY FUND). CUP: E77G23000150001

\bibliographystyle{unsrt}

\bibliography{Bibliography}

% If authors have biography, please use the format below
\end{document}